\documentclass[preprint, 12pt]{elsarticle}
\usepackage{lineno,hyperref}
\usepackage{amsmath}
\usepackage[romanian]{babel}
\usepackage{combelow}
\journal{Physics of the Dark Universe}

\begin{document}

\begin{frontmatter}

\title{The contribution of the quantum vacuum to the cosmological constant is zero: proof that vacuum energy does not gravitate}

\author[1]{G. B. Mainland\corref{cor1}}
%\cortext[mycorrespondingauthor]{G. B. Mainland}
\ead{mainland.1@osu.edu}

\author[2]{Bernard Mulligan}
\ead{mulligan.3@osu.edu}

\cortext[cor1]{Corresponding author}

\address[1]{Department of Physics, The Ohio State University at Newark, 1179 University Dr., Newark, OH 43055, USA}
\address[2]{Department of Physics, The Ohio State University, Columbus, OH 43210, USA}


%\cortext[mycorrespondingauthor]{G. B. Mainland}
%\ead{mainland.1@osu.edu}


\begin{abstract}
The consensus among many  theoretical physicists is that the calculated contribution of the quantum vacuum to the total energy density of the universe is approximately $10^{121}$ times the observed energy density.  This is thought to be one of the worst theoretical predictions of all time. However, as shown here, this immense vacuum energy cannot in and of itself exert forces on normal matter.  As a result the huge vacuum energy density predicted by quantum field theory does not contribute to the ordinary energy density of the universe, is not a source for gravitational fields, and, as a result, does not contribute to the value of the cosmological constant.
\end{abstract}

\begin{keyword}
cosmological constant; vacuum energy density; quantum vacuum; bound particle-antiparticle vacuum fluctuation; photon-antiphoton vacuum fluctuation;
\end{keyword}

\end{frontmatter}

%\linenumbers

\section{Introduction:  the quantum vacuum}
\label{sec:1}

The present consensus is that quantum field theory predicts a vacuum energy density that is much too large.  The Planck CMB anisotropy measurements \cite{Planck:18} determined that the energy density in the universe is, to within 0.4\%, equal to the critical energy density. However, in a standard calculation, if no cutoff is imposed, the  vacuum energy density resulting just from photons is infinite;  if the maximum energy of a photon is cut off at the Planck energy,  the contribution of the vacuum energy density of photons is approximately $10^{121}$ times the observed energy density of the universe.  (Including the contribution from gravitons and gluons increases the  value of the theoretical calculation by about an order of magnitude. The contribution from fundamental massive particles is less.) The huge disparity between the observed energy density of the universe and the calculation of the vacuum energy density is known as the ``vacuum catastrophe'' \cite{Weinberg:89,Adler:95}. 

Since it is certain experimentally that the gigantic vacuum energy is not present as normal energy, the theoretical response has generally been to ignore it: the contribution of the vacuum energy to the total energy density of the universe has been simply assumed to be small or zero although the theoretical mechanism is unknown.  Blau and Guth\cite{Hawking:87}, for example, write,``this theoretical estimate [of the vacuum energy density] is regarded by many to be one of the deepest mysteries of physics.''

The ``vacuum catastrophe''  is a long-standing problem.  In their 2002 review article, Rugh and Zinkernagel\cite{Rugh:02} summarized the situation with regard to the cosmological constant $\Lambda$ by focusing on the relation between quantum field theory and general relativity. Such a focus requires a resolution of why the vacuum energy density appears to give the cosmological constant a value vastly larger than the observed value.  Explaining why the effective cosmological constant is not large is known as ``the old cosmological constant problem''\cite{Wang:17,Perlmutter:99}, which is the topic  primarily discussed in this article.  By carefully examining the predictions of quantum field theory and employing well-known physics principles,  it is demonstrated here that the huge vacuum energy density predicted by quantum field theory cannot exert a gravitational force; the vacuum energy density makes no contribution to the  cosmological constant. 

As will be discussed, the manifestation of vacuum energy is through the appearance of vacuum fluctuations. Accordingly, the plan of this article is as follows: The structure and properties of vacuum fluctuations are presented in Sec.~2. In Sec.~3 quantum field theory is used to prove that vacuum fluctuations with their structure and accompanying properties must exist, and expressions are derived for the vacuum expectation value of the energy associated with these fluctuations.  In Sec.~4 expressions are derived for the energy density of bound, massive particle-antiparticle vacuum fluctuations and photon-antiphoton vacuum fluctuations, the latter of which, when a cutoff is imposed at the Planck energy, is shown to be approximately $10^{121}$ times the observed energy density of the universe.  

In Sec.~5  Einstein's field equation is used to show that the vacuum expectation value of the energy-momentum tensor has the form of a cosmological constant. The vacuum energy density of the universe is shown to be constant, implying that the vacuum energy density is conserved independently of the conservation of energy for normal matter.   As will be discussed in Sec.~6, a consequence of conservation of the vacuum energy density is that vacuum energy cannot exert forces on normal matter, either directly or indirectly through the presence of vacuum fluctuations. Therefore the vacuum energy density cannot be a source for gravitational fields\footnote{The restricted ways in which vacuum fluctuations can interact with normal matter are discussed in \cite{Mainland:20} and make it possible to calculate the permittivity and the speed of light in the vacuum.}.    Because vacuum energy  cannot exert a force on normal matter, it cannot contribute to the value of the cosmological constant. Thus the contribution to the cosmological constant from vacuum energy is zero. In Sec.~7 the results of the article are summarized and the consequences of those results are discussed.  

\section{Structure and properties of vacuum fluctuations}
  \label{sec:2}

As has been emphasized by Dimopoulos, Raby, and Wilczek\cite{Dimopoulos:91} and further developed by Wilczek\cite{Wilczek:08} in terms of his concept of the Grid, the quantum vacuum is filled with fluctuations that may interact with normal (meaning real, ordinary, and observable) test particles placed in the vacuum.  Physicists use the term ``vacuum fluctuation'' to describe two very different entities that have been designated ``type 1,'' which have observable consequences, and ``type 2,'' which have no observable consequences\cite{Mainland:20}. Type 2 vacuum fluctuations are  sometimes called vacuum diagrams\cite{Jauch:76} or vacuum bubbles\cite{Bjorken:65}, a class of Feynman diagrams for which virtual particles appear from and then vanish back into the vacuum. These diagrams, and as a consequence type 2 vacuum fluctuations,  do not contribute to physical processes.  Type 2 vacuum fluctuations consist of virtual particles that are off shell, appear as internal particles in Feynman diagrams, and exist for a time $\Delta t$ permitted by the Heisenberg uncertainty principle.

Clearly, the appearance of a type 1 vacuum fluctuation must obey the laws of physics: (a) Angular momentum must be conserved, implying that the vacuum fluctuation itself must have zero angular momentum. Additional conserved quantities are (b) electric charge, (c) lepton number, and (d) baryon number.  Individual leptons or baryons cannot appear as vacuum fluctuations because their appearance would violate both conservation of angular momentum and conservation of lepton or baryon number, respectively.  Type 1 vacuum fluctuations appear as particle-antiparticle pairs and possess two additional properties:  (e) Vacuum fluctuations are on shell.  The proof of the existence of vacuum fluctuations is given in Sec.~3. (f) In an inertial reference frame, vacuum fluctuations cannot exert forces on normal particles, as discussed in Sec.~6.  If the particle and antiparticle in the pair are massive, to minimize the violation of conservation of energy allowed by the Heisenberg uncertainty principle, the pair must appear with zero center-of-mass momentum in the least energetic bound state that has zero angular momentum.  In the remainder of this article the acronym ``VF'' will refer exclusively to a massive type 1, bound, particle-antiparticle vacuum fluctuation.  The acronym ``PVF'' will refer to a type 1, photon-antiphoton vacuum fluctuation.  Because a photon is its own antiparticle, a PVF consists of two, on-shell photons traveling in opposite directions, each with the same helicity, so that the total angular momentum is zero.  The  VF or PVF exists for a time $\Delta t$ permitted by the Heisenberg uncertainty principle.\cite{Heisenberg:83,Thirring:58,Bjorken:65}.

Normal particles can, under very restricted circumstances, be used as test particles to observe characteristics of the Grid. The calculation by the authors\cite{Mainland:19,Mainland:20} of the permittivity $\epsilon_0$ of the vacuum,  the speed  $c$ of light in the vacuum, and the fine structure constant $\alpha$ validates the statement  by Dimopoulos, Raby, and Wilczek\cite{Dimopoulos:91} that ``the vacuum is a dielectric,'' and establishes that the major contribution to the dielectric constant of the vacuum results from VFs of charged lepton-antilepton pairs that appear in the vacuum as bound states.  A formula for the permittivity $\epsilon_0$ of the vacuum was calculated by examining the interaction of photons with VFs that are bound, charged lepton-antilepton pairs.  Then a formula for $c$ was immediately obtained from the equation $c=1/\sqrt{\epsilon_0 \mu_0}$, where $\mu_0$ is the permeability of the vacuum. Using the formulas for $\epsilon_0$ and $c$, a value for the fine-structure constant was calculated\cite{Mainland:19} that, to lowest order in $\alpha$, agrees with the experimental value to within a few percent. 

The calculation of $c$ must satisfy three conditions, one from special relativity and two from electrodynamics. In Einstein's 1905 paper\cite{Einstein:05} ``On the electrodynamics  of moving bodies'' in which he introduced special relativity, one of the two postulates on which his work is based is that the speed of light in the vacuum is the same in every inertial reference frame.  Furthermore, Einstein was compelled to introduce the assumption that there is no preferred inertial frame of reference, as demonstrated by Michelson and Morley in 1887\cite{Michelson:87} and by experiments that continued to search for such a frame\cite{Ohanian:08}.   As Leonhardt et al.\cite{Leonhardt:18} state,``In free space the vacuum is Lorentz invariant, so a uniformly moving observer would not see any effect due to motion$\dots$.'' Since the vacuum is at rest with respect to every inertial reference frame and the speed of light in the vacuum is determined by the interaction of  photons with VFs\cite{Mainland:19,Mainland:20},  it follows  that the speed of light is the same in every inertial frame, providing a theoretical explanation for why Einstein's postulate is true and obviating the need for the postulate.  To make the equations of physics the same in all inertial frames of reference required modification of the equations describing the motion of particles, but not the field equations of electricity and magnetism\footnote{In this article, as in references \cite{Mainland:20,Mainland:19}, SI units are used throughout.  The present article deals with fundamental issues that need to be understood by nonspecialists. The specialist who finds the explicit use of S.I. units a nuisance does not need the presence of $\hbar$ and $c$ for clarification.}.
.

For the derivations of $\epsilon_0$ and $c$ to be consistent with Maxwell's equations, the derivations must satisfy the following two general relations when restricted to the vacuum:  The electric displacement $\mathbf{D}$ in a dielectric satisfies $\mathbf{D}=\epsilon_0 \mathbf{E}+\mathbf{P}$, where  $\mathbf{E}$ and $\mathbf{P}$ are, respectively,  the electric field and  the  polarization density.  In the vacuum this relation becomes $\mathbf{D}=\epsilon_0 \mathbf{E}$, a relation that was used  in the derivation of the formula for $\epsilon_0$.  The second electrodynamics relation that must be satisfied follows from the general equation $\mathbf{B}=\mu_0 (\mathbf{H}+\mathbf{M})$, where $\mathbf{B}, \mathbf{H}$ and $\mathbf{M}$ are, respectively, the magnetic field, the magnetic field strength, and the magnetization. In the vacuum the relation becomes $\mathbf{B}=\mu_0 \mathbf{H}$, which the derivations satisfy  because bound, charged lepton-antilepton  vacuum fluctuations have zero total angular momentum and, therefore, have zero magnetism $\mathbf{M}$\cite{Sauder:67}.

\section{The creation and zero-point energy of vacuum fluctuations}
\label{sec:3}
 
Vacuum fluctuations result from fluctuations of free fields so they are on shell. As discussed in the previous section, they always consist of particle-antiparticle pairs to help ensure that all conservation laws, including conservation of angular momentum, are satisfied.  The pair must also minimize the violation of conservation of energy allowed by the Heisenberg uncertainty principle.  As a result, in an inertial reference frame a massive, bound, particle-antiparticle pair appears on shell, has zero angular momentum and zero center-of-mass momentum. 

The center of mass of a bound,  particle-antiparticle VF is stationary while the particle-antiparticle spin-0 pair undergoes zitterbewegung (``trembling motion'') with an amplitude\cite{Broglie:52,Barut:81} 
\begin{equation}\label{eqn:1}
\frac{\hbar}{2Mc}\equiv L_{\rm Z}
\end{equation}
that gives the VF a size.  In the above equation $M$ is the mass of the bound, massive particle-antiparticle VF. The volume $\mathbf{V}_{\rm Z}  \equiv  L_{\rm Z}^3$ resulting from zitterbewegung is the  average volume in which one VF exists.  The amplitude $L_{\rm Z}$ associated with the zitterbewegung depends only on the mass of the particle, not its spin\cite{Corinaldesi:63}. A VF or PVF must have spin-0\footnote{In the case of a particle with spin $\neq 0$, the spin of the particle is expressed by a rotation of the particle around an axis of orientation of the particle.  Whether the particle is spinning or not, the amplitude of the zitterbewegung is that stated in \eqref{eqn:1}\cite{Corinaldesi:63}.}.

Because the quanta of a free field behave as free particles, to understand the creation of a VF or a PVF it is necessary to represent a VF or PVF by a quantum field. Field theory provides a proof that VFs exist and a formula for the source of the energy available for the creation of  VFs.  The structure of a  VF is not important when discussing  the creation of  the VF. What is most important is that the VF's total angular momentum is zero.  Thus  a  VF in its  ground state with zero angular momentum can be approximately represented  by the Klein-Gordon field $\phi(x)$ for a free, neutral,  spin-0 particle\cite{Pauli:94,Wentzel:03,Thirring:58}. To show that VFs  must exist, first note that the  free field $\phi(x)$ contains two terms, one proportional to a creation operator $a^\dagger_\mathbf{k}$ and the other proportional to an annihilation operator $a_\mathbf{k}$. The vacuum expectation value of each operator is zero, so the average value of a free field in the vacuum is zero,
\begin{equation}\label{eqn:2}
(0|\phi(x)|0)=0 \,. 
\end{equation}
The expectation value  of the product of a free field at two different locations $x$ and $x^\prime$ is written in terms of the angular wave number $\mathbf{k}=\mathbf{p}/\hbar$ and the angular frequency $\omega_\mathbf{k}=E/\hbar$ where $\mathbf{p}$ and $E$ are, respectively, the relativistic momentum and energy\cite{Thirring:58,Bjorken:65},
\begin{equation}\label{eqn:3}
(0|\phi(x)\phi(x^\prime)|0)= \hbar \int_{k_0=\omega_\mathbf{k}/c}  \frac{{\rm d}^3k}{(2\pi)^32\omega_\mathbf{k}}e^{-ik^\mu(x_\mu-x^\prime_\mu)}\,.
\end{equation}

The expression in  \eqref{eqn:3}. is nonzero because the product $\phi(x)\phi(x^\prime)$ contains a term proportional to $a_\mathbf{k}\,a^\dagger_\mathbf{k^\prime}$ that has a nonzero vacuum expectation value\cite{Bohm:19}.  Eq. \eqref{eqn:3} has the feature that $(0| \phi^2(x) |0)$ is infinite.   However, as a result of zitterbewegung, any VF  has a  finite  size $\mathbf{V}_{Z}$ over which it must be averaged, thus removing the infinity in $(0|\phi^2(x)|0)$. Because the vacuum expectation value of the product $\phi(x)\phi(x^\prime)$ is nonzero, the  free field $\phi(x)$ in the vacuum cannot be zero everywhere although its average value given in \eqref{eqn:2} is zero.
 
The  Hamiltonian $H$ of the Klein-Gordon field\cite{Wentzel:03,Bjorken:65} is
\begin{equation}\label{eqn:4}
H=\sum_\mathbf{k} \,\hbar\,\omega_\mathbf{k} \left( a^\dagger_\mathbf{k}\,a_\mathbf{k} +\frac{1}{2}\right),\hspace{0.3 cm}\omega_\mathbf{k}=+\sqrt{c^2\mathbf{k}^2+\frac{M^2c^4}{\hbar^2}}  \,.
\end{equation}
The above Hamiltonian has the same form as that of a harmonic oscillator.   Since the annihilation operator acting on the vacuum is zero, $a_\mathbf{k}|0)=0$, the expectation value of the Hamiltonian in the vacuum is 
\begin{equation}\label{eqn:5}
(0|H|0)=\sum_\mathbf{k} (0|\hbar\,\omega_\mathbf{k}\left( a^\dagger_\mathbf{k}\,a_\mathbf{k} +\frac{1}{2}\right)|0)= \frac{1}{2}\sum_\mathbf{k}\hbar \, \omega_\mathbf{k} \,.
\end{equation}
 The energy in the vacuum, the zero-point energy, is the term on the right-hand side of \eqref{eqn:5}. Because there are an infinite number of cells in $\mathbf{k}$-space,  the energy in the vacuum is infinite. However, in any inertial frame a VF must appear in the vacuum at rest so its center-of-mass momentum  $\mathbf{p}=\hbar \mathbf{k}=0.$ In \eqref{eqn:5} the only energy available to create a VF is the finite energy $\hbar  \omega_\mathbf{k=0}/2$.  The factor of 1/2 is present  in   \eqref{eqn:5} because on average the VF is present half of the time.
\begin{equation}\label{eqn:6}
(0|H|0)_{VF}= \frac{1}{2}\hbar \, \omega_\mathbf{k=0}=\frac{1}{2}Mc^2 \,.
\end{equation}

Since VFs are represented by noninteracting fields, the quanta associated with the fields are on shell: from \eqref{eqn:3} it follows that the integral over three-dimensional $\mathbf{k}$-space satisfies the condition $k_0=\omega_{\mathbf{k}}/c$. Using the expression for  $\omega_k$ in  \eqref{eqn:4}, it  is easy to verify that VFs satisfy the on-shell condition $E^2-(\mathbf{p}c)^2= (\hbar \omega_\mathbf{k})^2-( \hbar \mathbf{k} c)^2= (\hbar c k_0)^2-( \hbar c \mathbf{k})^2=  (\hbar c)^2k^\mu k_\mu=(Mc^2)^2$.
 
Since the mass of a photon is zero, polarization and helicity are measures of the same quantity: a single photon can be represented either by positive or negative helicity. Angular momentum is conserved when a PVF appears in the vacuum with the two photons moving in opposite directions, each with the same helicity.  The zero-point energy of a PVF is then given by the right-hand side of \eqref{eqn:5}\cite{Lifshitz:82}  after the mass $M$  is set to zero and the two helicity states are summed over. Since there are two photons in a  PVF,
\begin{equation}\label{eqn:7}
(0|H|0)_{\rm PVF}= (2\;\mbox{photons})(2\;\mbox{helicities})\frac{1}{2}\sum_\mathbf{k} \hbar \,\omega_\mathbf{k}\big|_{M=0}=2 \sum_\mathbf{k}\hbar c|\mathbf{k}| \,.
\end{equation}
The extension of the 1911 Planck result for photons\cite{Planck:11}  to include all quantum fields  goes at least as far back as Marshall\cite{Marshall:63}.

\section{Energy density of VFs and PVFs and their interaction with normal matter}
 \label{sec:4}
  
 In an inertial frame of reference, the number density of VFs, which appear  as bound states, is $1/L_Z^3$\cite{Mainland:20}.     The energy density is then the energy of a  VF multiplied by the number density,
\begin{equation}\label{eqn:8}
\rho^{\rm energy}_{VF}=\frac{\frac{1}{2}Mc^2}{L_z^3}=4\frac{(Mc^2)^4}{(\hbar c)^3}\,.
\end{equation}
 To obtain the total energy density resulting from all massive particle-antiparticle VFs,  \eqref{eqn:8}  must be summed over all fundamental, massive particles that form spin-0, particle-antiparticle bound states.  For each different particle the mass $M$ in \eqref{eqn:8} is the mass corresponding to  the energy of the state with spin-0 and minimum bound-state energy.

 The charged fermion-antifermion VF  that makes the smallest contribution to this sum is the electron-positron VF that appears in the vacuum as parapositronium. Then $M=2m_e -\mbox{binding energy}$, where $m_e$ is the mass of the electron. Neglecting the binding energy, which is very small in comparison with $2m_e$,
\begin{equation}\label{eqn:9}
\rho^{\rm energy}_{\rm parapositronium VF}\cong9.1\times 10^{25}{\rm J/m^3}\,.
\end{equation}
The experimental value of the energy density of the universe\cite{Tanabashi:18} is
\begin{equation}\label{eqn:10}
\rho^{\rm energy}_{\rm universe}=7.8\times 10^{-10}{\rm J/m^3}\,.
\end{equation}
Thus the energy density resulting from parapositronium VFs  alone is $10^{35}$ times the observed energy density of the universe.

The energy density of photon-antiphoton vacuum fluctuations is calculated as follows\cite{Weinberg:89,Adler:95}:  In \eqref{eqn:7} the entire $\mathbf{k}$-space is taken to be a cube with volume $L^3$, and that space is divided into cells labeled by integers. Periodic boundary conditions\cite{Saxon:68,Adler:95,Gasiorowicz:03} are imposed: along the x-axis the boundary condition is
\begin{equation}\label{eqn:11}
e^{ik_x L}=1\;\;\mbox{or}\;\; k_xL=2\pi n_x\,,
\end{equation}
where $n_x$ is a positive or negative integer.   Corresponding relations are obtained for the y- and z-axes:
\begin{equation}\label{eqn:12}
(n_x, n_y, n_z)=\frac{L}{2\pi}(k_x, k_y, k_z)\,.
\end{equation}

From \eqref{eqn:7} the energy density of PVFs is
\begin{equation}\label{eqn:13}
\rho^{\rm energy}_{\rm PVF}=\frac{1}{L^3} \sum_\mathbf{k}2 \hbar c|\mathbf{k}| \,.\end{equation}
The sum over $\mathbf{k}$ in discrete $\mathbf{k}$-space becomes a sum over integers that, in turn, can be expressed as an integral:
\begin{equation}\label{eqn:14}
\rho^{\rm energy}_{\rm PVF}=\frac{2 \hbar c}{L^3} \sum_{n_x}\sum_{n_y}\sum_{n_z} |\mathbf{k}| \rightarrow \frac{2 \hbar c}{L^3}  \int {\rm d}^3n\; |\mathbf{k}|  \,.      
\end{equation}
From \eqref{eqn:12}, 
\begin{equation}\label{eqn:15}
{\rm d}^3n=\left( \frac{L}{2 \pi} \right)^3 {\rm d}^3k \,.
\end{equation}
Using \eqref{eqn:15}, \eqref{eqn:14} becomes
\begin{equation}\label{eqn:16}
\rho^{\rm energy}_{\rm PVF}= \frac{2 \hbar c}{(2\pi)^3}  \int {\rm d}^3k\; |\mathbf{k}|  \,.
\end{equation}
Transforming from Cartesian to spherical coordinates and noting that the integrand is spherically symmetric, the angular integration can be performed immediately, yielding a factor of $4\pi$,
\begin{equation}\label{eqn:17}
\rho^{\rm energy}_{\rm PVF}= \frac{4 \hbar c}{(2\pi)^2}  \int_0^\infty {\rm d}|\mathbf{k}|\; |\mathbf{k}|^3 \,.
\end{equation}
Imposing a cutoff $|\mathbf{k}|_{\rm{max}}$ for the divergent integral,
\begin{equation}\label{eqn:18}
\rho^{\rm energy}_{\rm PVF}= \frac{ \hbar c}{(2\pi)^2}  |\mathbf{k}|_{\rm{max}}^4 \,.
\end{equation}

 To obtain the total energy density resulting from all massless particle-antiparticle VFs,  \eqref{eqn:18}  must be summed over all fundamental, massless particles. In addition to photons, the vacuum energy of gravitons and the eight gluons must be included.  This will increase \eqref{eqn:18} by about an order of magnitude, which will not change any conclusions, so the effects of gluons and gravitons will not be included.

Choosing the cutoff so that the maximum energy of a photon is the Planck energy $E_P=\sqrt{\hbar c^5/G}$, where $G$ is the gravitational coupling constant, $|\mathbf{k}|_{\rm{max}}=E_P/(\hbar c)=6.2\times 10^{34}\,$m$^{-1}$. Eq. \eqref{eqn:18} then becomes
\begin{equation}\label{eqn:19}
\rho^{\rm energy}_{\rm PVF}=1.2 \times 10^{112}\,\mbox{J/m}^3 \,.
\end{equation}
The energy density resulting from PVFs   is $10^{121}$ times the observed energy density of the universe, the result stated in the Abstract.
 
 \section{The effective cosmological constant}
\label{sec:5}

Einstein's field equation\cite{Weinberg:89,Rugh:02,Martin:12,Wang:17} with a bare cosmological constant $\Lambda_{\rm bare}$  is
\begin{equation}\label{eqn:20}
R_{\mu\nu}-\frac{1}{2}R\, g_{\mu\nu}-\Lambda_{\rm bare}\, g_{\mu\nu}=-\frac{8\pi G}{c^4}T_{\mu\nu}\,.
\end{equation}
In \eqref{eqn:20} $R_{\mu\nu}$  is the Ricci tensor, $R \equiv R^\sigma_\sigma$ is the Ricci scalar,  $g_{\mu\nu}$ is the metric tensor, $G$ is the gravitational force constant, and   $T_{\mu\nu}$ is the energy-momentum tensor.  As will be shown, the vacuum expectation value of $T_{\mu \nu}$, namely $(0|T_{\mu \nu}|0)$, is the part of $T_{\mu \nu}$ that has the mathematical form of a cosmological constant in \eqref{eqn:20}\cite{Carroll:92}.  To isolate the contribution from this term,
the energy-momentum tensor resulting entirely from the presence of normal matter is, by definition,
\begin{equation}\label{eqn:21}
T_{\mu \nu}^{\rm matter} \equiv T_{\mu\nu}-(0|T_{\mu\nu}|0)\,,
\end{equation}
because it has an expectation value of zero in the vacuum.  In terms of $T_{\mu\nu}^{\rm matter}$, \eqref{eqn:20} becomes 
\begin{equation}\label{eqn:22}
R_{\mu\nu}-\frac{1}{2}R\, g_{\mu\nu}-\Lambda_{\rm bare}\, g_{\mu\nu}=-\frac{8\pi G}{c^4}[T_{\mu\nu}^{\rm matter}+(0|T_{\mu\nu}|0)]\,.
\end{equation}

To show that  the vacuum expectation value of $T_{\mu\nu}$  has the form of a cosmological constant\cite{Carroll:92}, $(0|T_{\mu\nu}|0)$ will first be considered in Minkowski space where $(0|T_{\mu\nu}|0)$ is a  $4 \times 4$ tensor.  Since the vacuum is the same in any inertial reference system,  $(0|T_{\mu\nu}|0)$ must be the same in any inertial reference frame. The only $4 \times 4$  tensor in Minkowski space that is invariant under  Lorentz boosts is the (diagonal) metric tensor $\eta_{\mu\nu}$, chosen here with diagonal elements (-1,+1,+1,+1); therefore, $(0| T_{\mu\nu} |0)$ must be proportional to $\eta_{\mu\nu}$, and the proportionality constant must be a scalar, denoted by $\cal{S}$,
\begin{equation}\label{eqn:23}
(0| T_{\mu\nu} |0)={\cal S} \eta_{\mu\nu}\,.
\end{equation}

The vacuum cannot conduct heat and has no shear stresses or viscosity, which are precisely the properties of a perfect fluid; therefore, the energy-momentum tensor of the vacuum is that of a perfect fluid\cite{Misner:73}.
\begin{equation}\label{eqn:24}
 T_{\mu\nu} =\left(\rho^{\rm mass}+\frac{P}{c^2}\right )U_\mu U_\nu +P\, \eta_{\mu\nu}\,,
\end{equation}
where  $\rho^{\rm mass}$ is the mass density,  $P$ is the isotropic pressure, and $U_\mu$ is a 4-velocity vector.  Since the vacuum is at rest with respect to every inertial frame, in an inertial frame the three-velocity $\mathbf{v}$ of the perfect fluid is zero. Using $U_\mu(\mathbf{v}=0)=(c,0,0,0)$ and the fact that the energy density  $\rho^{\rm energy}=\rho^{\rm mass}c^2$, a perfect fluid at rest has the diagonal energy-momentum tensor $T_{\mu,\nu}$ where
\begin{equation}\label{eqn:25}
T_{00}=\rho^{\rm energy},\;T_{11}=T_{22}=T_{33}=P\,.
\end{equation}
Defining $\rho^{\rm energy}_{\rm vac}\equiv(0|\rho^{\rm energy}|0)$ and $P_{\rm vac}\equiv (0|P|0)$ and then substituting \eqref{eqn:25} into \eqref{eqn:23} establishes that  \eqref{eqn:23} is satisfied provided
\begin{subequations}\label{eqn:26}
\begin{align}
\label{eqn:26a}
&{\cal S}= -\rho^{\rm energy}_{\rm vac}\,,\\
\label{eqn:26b}
&P_{\rm vac}=-\rho^{\rm energy}_{\rm vac}\,.
\end{align}
\end{subequations}
Using \eqref{eqn:26a},  \eqref{eqn:23} becomes
\begin{equation}\label{eqn:27}
(0| T_{\mu\nu} |0)=-\rho^{\rm energy}_{\rm vac} \eta_{\mu\nu}\,.
\end{equation}
The generalization of  \eqref{eqn:27} to curved space-time is 
\begin{equation}\label{eqn:28}
(0|T_{\mu\nu}|0)=-\rho^{\rm energy}_{\rm vac}\; g_{\mu\nu}\,.
\end{equation}
Using  \eqref{eqn:28}, \eqref{eqn:22} becomes  
\begin{equation}\label{eqn:29}
R_{\mu\nu}-\frac{1}{2}R\, g_{\mu\nu}-(\Lambda_{\rm bare}+\frac{8\pi G}{c^4}\rho^{\rm energy}_{\rm vac})g_{\mu\nu}=-\frac{8\pi G}{c^4}T_{\mu\nu}^{\rm matter}\,,
\end{equation}
from which it follows that the effective cosmological constant $\Lambda_{\rm eff}$ is
\begin{equation}\label{eqn:30}
\Lambda_{\rm eff}= \Lambda_{\rm bare}+ \frac{8\pi G}{c^4}\rho^{\rm  energy}_{\rm  vac}\,.
\end{equation}

As a result of  \eqref{eqn:26b}, which is the equation of state for the vacuum, the vacuum energy density remains constant as the universe expands (or contracts) adiabatically\cite{Carroll:92,Carroll:01,Wilczek:08}. As a result the vacuum energy density is neither a function of time nor position: $\rho^{\rm energy}_{\rm vac}$ is a constant, independent of time and uniform throughout all space. Since vacuum energy density is conserved, vacuum energy is conserved independently of ordinary energy in the physical world;  energy conservation in the physical world is well established without regard to vacuum energy.

Eq. \eqref{eqn:30} is the source of ``the old cosmological constant problem.''  For  an order-of-magnitude estimate of the final term in \eqref{eqn:30}, if $\rho^{\rm  energy}_{\rm  vac}$ is approximated (but underestimated) by the vacuum energy density of photons as given in \eqref{eqn:19},
\begin{equation}\label{eqn:31}
 \frac{8\pi G}{c^4}\rho^{\rm  energy}_{\rm  vac} \sim \frac{8\pi G}{c^4}\rho^{\rm  energy}_{\rm PVF}  \sim 10^{69}{\rm m}^{-2}\,.
\end{equation}
The experimental value of the cosmological constant follows from $\Omega_\Lambda=0.692$\cite{Tanabashi:18} and the critical density $\rho_{\rm  critical}=8.66\times 10^{-27}$ kg/m$^3$\cite{Tanabashi:18} of the universe: $\Lambda_{\rm eff} = 1.12 \times10^{-52} {\rm m}^{-2}$,  which is approximately $10^{-121}$ times smaller than the final term on the right-hand-side of  \eqref{eqn:31}. This is ``the old cosmological problem''. 

\section{Why the contribution of the quantum vacuum to the cosmological constant is zero}
\label{sec:6}

Because the vacuum energy density is conserved,  VFs and PVFs  can interact with normal matter only  in specific, limited ways.  As an example,  a normal photon traveling through the vacuum can interact  with a VF to create a photon-excited VF. The progress of the photon through the vacuum is therefore slowed. When the VF annihilates, the energy borrowed from the vacuum for its creation must be returned; therefore,  a photon identical to the original photon is emitted and continues through the vacuum.  This interaction determines the speed of light in the vacuum\cite{Mainland:19,Mainland:20}.  

There is a second way that VFs can interact with normal matter:  consider a VF that occurs for a  specific type of particle-antiparticle pair such as an electron-positron pair.  A normal electron could annihilate with the positron that was part of the VF, returning to the vacuum the energy originally borrowed to create the VF.  The electron that was part of the VF would then become a normal electron at a location different from the original, normal electron, giving rise to zitterbewegung\cite{Thirring:58}.

A solution to ``the old cosmological constant problem'' thus follows immediately from an understanding of vacuum fluctuations. Eq.  \eqref{eqn:30} appears  to show that  the bare cosmological constant $\Lambda_{\rm bare}$ is increased by an amount  $8\pi G \rho^{\rm  energy}_{\rm  vac}/c^4$ as a result of the presence of the vacuum energy density $\rho^{\rm  energy}_{\rm  vac}$.  Appearances are deceiving.

So why doesn't the vacuum energy density contribute to the cosmological constant?  The presence of the term proportional to $\rho^{\rm  energy}_{\rm  vac}$ in  \eqref{eqn:30} depends on the assumption that the vacuum energy density can exert a classical gravitational force on a normal particle. However, in an inertial frame the vacuum energy density {\bf cannot}. If vacuum energy could, it would do work on normal particles, permanently transferring energy between the vacuum and the physical world and reducing the vacuum energy density.  

Since vacuum fluctuations are the quantum manifestation of vacuum energy, if vacuum energy could exert a gravitational force, that force would result from the interaction of normal matter with a vacuum fluctuation.  When the vacuum fluctuation vanished back into the vacuum, the energy associated with any work done by the vacuum fluctuation would be remain behind as ordinary energy, permanently decreasing the vacuum energy density and violating the separate conservation of energy in the physical world and energy density in the vacuum.  From a quantum perspective, if a VF has not interacted with normal matter and gained energy, it cannot spontaneously emit quanta unless it reabsorbs identical quanta: all it can do is vanish back into the vacuum, returning to the vacuum the energy borrowed for its creation.  If a VF has formed a quasi-stationary state with a normal boson such as a photon or graviton, when the VF vanished back into the vacuum,  to conserve energy, momentum, and angular momentum, all the VF can do is emit a boson that is identical to the original.  As was shown by the authors\cite{Mainland:19,Mainland:20}, this is precisely the mechanism by which photons (and almost certainly gravitons\cite{Ford:77}) travel through the vacuum. Since a VF cannot ``permanently'' exchange a boson associated with a force with normal matter, it cannot exert a force on normal matter.  

From the conservation of the energy density, an observer in an inertial reference frame will conclude that vacuum energy cannot exert a force on normal matter so the vacuum energy of the vacuum does not contribute to the energy density of normal energy in the universe, explaining why there is no ``vacuum catastrophe''.   Similarly, in an inertial frame vacuum energy cannot exert a gravitational force on normal matter so vacuum energy cannot be a source for a gravitational field.  Thus in \eqref{eqn:30} $\rho^{\rm  energy}_{\rm  vac}$ contributes nothing to the cosmological constant and $\Lambda_{\rm eff}= \Lambda_{\rm bare}$. That is, quantum vacuum energy does not gravitate.
 
\section{Results and discussion}
\label{sec:7}

VFs are on shell, massive particle-antiparticle pairs.  In an inertial frame they appear at rest in the least energetic bound state that has zero angular momentum. PVFs are photon-antiphoton pairs. In an inertial frame the two photons appear with zero center-of-mass momentum and the same helicity so that each pair has zero total angular momentum.  Field theory predicts the existence of both VFs and PVFs as well as their energy in the vacuum.  

The vacuum energy density of VFs and PVFs is both time-independent and homogeneous so in an inertial frame of reference, vacuum energy is conserved independently of normal energy.  A consequence is that in an inertial frame, vacuum energy cannot exert a force on normal matter. The calculations presented in this paper directly address  the issue raised by Wang, Zhu, and Unruh\cite{Wang:17} with respect to the combination of quantum field theory and general relativity:  ``the equivalence principle of general relativity requires that every form of energy gravitates in the same way.''  The conclusion of the present article is that in an inertial frame of reference vacuum energy cannot exert a gravitational force because exerting such a force would  violate conservation of vacuum energy density.  Accordingly, in an inertial frame, vacuum energy does not gravitate.  Therefore, in an inertial frame vacuum energy density does not contribute to the  normal energy density in the universe, explaining why there is no ``vacuum catastrophe''.  

Physicists have struggled to explain why the cosmological constant is so small. Before the cosmological  constant was found to be nonzero\cite{Riess:98,Perlmutter:99},  Barrow and Shaw\cite{Barrow:11} wrote,``many particle physicists suspected that some fundamental principle must force [the value of the cosmological constant] to be precisely zero.Ó Perlmutter et al.\cite{Perlmutter:99} wrote,``$\dots$ presumably some symmetry of the particle physics model is causing cancellations of this vacuum energy density.'' Since  vacuum energy cannot exert a gravitational force in a inertial frame, vacuum energy is not a source for a gravitational field and does not contribute to the value of the cosmological constant, solving ``the old cosmological constant problem.''

Determining why the cosmological constant has the observed value is very much an open question\cite{Copeland:06,Brax:18}.  As Guth\cite{Guth:81} wrote in 1981,``The reason $\Lambda$ is so small is of course one of the deep mysteries of physics. The value of $\Lambda$ is not determined by the particle theory alone, but must be fixed by whatever theory couples particles to quantum gravity.'' 

%\externalbibliography{yes}
\bibliography{CosConst}
\end{document}